\documentclass[twocolumn,a4paper]{article-nd2001_2}
\usepackage{times,nd2001_2}
\usepackage[dvips]{graphicx}
\usepackage{jnstcite}
\begin{document}
\title{\bf {\it s}-Process Nucleosynthesis in Low-Metallicity Stars}  
\author{ Nobuyuki IWAMOTO$^{1,\ast}$, Toshitaka KAJINO$^{1,2}$,
         Grant J. MATHEWS$^{3}$, \\ Masayuki Y. FUJIMOTO$^{4}$ 
         and Wako AOKI$^{1}$ \\
\begin{tabular}{c}
\vspace{-3.5mm}
 \\
\fontsize{9}{12}
  $^{1}$\textit{National Astronomical Observatory, 2-21-1 Osawa, Mitaka,
 Tokyo 181-8588, Japan} \\
\fontsize{9}{12}
  $^{2}$\textit{The Graduate University for Advanced Studies, 2-21-1
 Osawa, Mitaka, Tokyo 181-8588, Japan} \\
\fontsize{9}{12}
  $^{3}$\textit{Center for Astrophysics, Department of Physics,
 University of Notre Dame, Notre Dame, IN 46556, USA} \\
\fontsize{9}{12}
  $^{4}$\textit{Department of Physics, Hokkaido University, Kita 10-jo
 Nisi 8-tyome, Sapporo, Hokkaido 060-0810, Japan} \\
\end{tabular}
}

\begin{abstract}
We have made a parametric study of {\it s}-process nucleosynthesis in the
metal poor ([Fe/H] $=-2.7$) stars LP625-44 and
LP706-7.  We find that a high neutron
exposure and a small overlap factor are necessary to fit the abundance
 pattern observed in these two metal-deficient 
stars, particularly the abundance ratios, Pb/Ba $\approx 1$ and
 Ba/Sr $\approx 10$.  We have also constructed stellar models
to better understand how such $s$-process conditions could arise.
We have calculated a 2$M_\odot$ model star with metallicity
 [Fe/H] $=-2.7$ from the ZAMS up to AGB phase. 
We find that for such low-metallicity stars 
the  He convective shell 
reaches the bottom of the overlying H-rich envelope at the second
 thermal pulse. Protons are then carried into the  hotter He burning layers
and  $^{13}$C is formed 
as protons mix into the He shell. 
Subsequently, material in the H-flash driven convective zone experiences a 
high neutron exposure due to the $^{13}$C($\alpha$, n) reaction. 
This results in a new neutron-capture {\it s}-process  paradigm in which
the abundances are characterized by only one neutron exposure.  
We suggest that this new {\it s}-process site may be a 
significant contributor to the
{\it s}-process abundances in low-metallicity ([Fe/H] $\le -2.5$) stars.
\begin{center} \begin{minipage}{145mm}
\end{minipage}\end{center}

\end{abstract}
\date{}

\maketitle

\keyword[~$^{\ast }$]{Corresponding author, Tel. +81-422-34-3735, Fax. +81-
422-34}
\keyword[~]{-3746, E-mail: iwamoto@th.nao.ac.jp}

\section{Introduction}

One of the important mechanisms by which heavy elements 
are produced in Nature is the slow
neutron capture process ({\it s}-process). In the {\it s}-process,  
the typical neutron capture timescale is longer than the beta-decay
timescale. Thus, heavy element synthesis from iron peak nuclei proceeds
along beta stability line and finally reaches lead (Pb) and bismuth. 

An extensive survey of metal deficient stars in the Galactic Halo
has recently been completed~\cite{Beers92}. One of the most exciting
discoveries to emerge from 
this survey is that many stars show enhancements of neutron
capture element~\cite{Hill00,Norris97}. Of particular relevance to 
this paper is a recently reported detailed
abundance analysis of LP625-44 and LP706-7 by Aoki {\it et al.}~\cite{Aoki00,Aoki01}.
These stars have almost identical metallicity as [Fe/H] $\sim -2.7$. 
However, unlike most metal
deficient stars, these two stars appear to have an almost
pure {\it s}-process abundance distribution. 
The LP625-44 star is observed to be in a binary system. Thus,
this star probably experiences mass transfer from 
a companion star
during its asymptotic giant branch (AGB) phase. By now, the companion star
has evolved into a 
white dwarf and can not be observed directly.

Regarding the $s$-process, it is of particular interest
to study the $s$-process abundances in metal-poor stars.
It is now commonly accepted that $^{13}$C($\alpha$, n)$^{16}$O is the
dominant 
major neutron source reaction in low mass AGB stars.
This reaction operates at temperatures greater than $9\times 10^{7}$K. 
Therefore, $^{13}$C burns radiatively in a narrow region at the top of 
the He intershell layer during a long interpulse phase near $\sim
10^4-10^5$~yr~\cite{Straniero95}. 

Naively, one expects that the $^{13}$C abundance produced in a star
should be independent of its metallicity.
This is because $^{13}$C 
is produced by proton capture on newly synthesized
$^{12}$C.
Furthermore, since the 
abundance of seed nuclei is low, it is
expected that these stars should show an abundance distribution
characterized by many neutron captures per seed.
Accordingly, theoretical model calculations 
made before the observations of these two stars
predicted  Pb/Ba $> 100$ and 
Ba/Sr $\sim 1$ at [Fe/H] $=-2.7$~\cite{Busso99}. 

LP625-44 and LP706-7 show, however, a peculiar distribution of {\it s}-process 
elements.  In particular,
 Pb/Ba $\sim 1$ and Ba/Sr $\sim 10$~\cite{Aoki00,Aoki01}.
These large differences between observation and theoretical prediction
may indicate that the most popular current 
interpretation~\cite{Busso99} of the nature
of the $s$-process in metal-deficient AGB stars
is not enough. Although Ryan {\it et al.}~\cite{Ryan00} found a good fit
to [Pb/Fe] and 
[Ba/Fe] for LP625-44, it was necessary to postulate an exceedingly low 
$^{13}$C abundance in order to reproduce the observation.  This 
eliminates any predictive power of the model. 

On the other hand, Fujimoto {\it et al.}~\cite{Fujimoto00} have proposed
that a new $s$-process paradigm 
may be at work in these low-metallicity AGB stars.
In that work it was found that the He convective shell which developed
during a thermal pulse could mix protons from the H-rich layer
for stars with   $M \le 3M_\odot$ and [Fe/H]
$\le -2.5$. This mixing event allows the production of 
$^{13}$C in the He convective shell
via the $^{12}$C($p$, $\gamma$)$^{13}$N($\beta^+\nu$)$^{13}$C 
reaction chain.

Quite recently, Van Eck {\it et al.}~\cite{vaneck01} reported
on abundances in three new
metal-deficient stars ([Fe/H] $=-2.45, -1.7,$ and $-1.65$).
These stars, indeed,  show an enhanced 
Pb abundance in comparison with other heavy elements.
This is consistent with the high neutron-to-seed ratio
expected in the standard theoretical model with partial mixing
from H-rich convective envelope~\cite{Mowlavi00}.

However, the metallicities of these three stars are larger than
[Fe/H] $=-2.5$, which is the critical value for the occurrence of the
Fujimoto {\it et al.}~\cite{Fujimoto00} $s$-process paradigm.
Thus, the Van Eck {\it et al.}~\cite{vaneck01} AGB stars may have 
evolved differently from two
metal-deficient stars, LP625-44 and LP706-7.
The purpose of this paper is to investigate the physical environment of
the {\it s}-process in these two peculiar metal-deficient AGB stars.

\section{Results}

\subsection{Parametric Studies of {\it s}-Process Nucleosynthesis}

We calculated {\it s}-process nucleosynthesis, using a schematic
pulsed-neutron-source model~\cite{Howard86} with updated neutron-capture reaction
rates~\cite{Bao00}. We assumed a fixed temperature of $10^8$~K
 during each neutron irradiation, 
consistent with
$^{13}$C($\alpha$, n)$^{16}$O as the dominant reaction for the neutron production. 
We characterize each pulse with a constant average
neutron number density ($N_n$) and a neutron exposure duration
($\Delta t$). The neutron exposure per pulse $\tau$ is defined by
\begin{equation}
\tau=\int N_n v_T dt \approx N_n v_T \Delta t,
\end{equation}
where $v_T$ is the average neutron thermal velocity.  The 
initial abundances of seed nuclei lighter than Fe group elements are
taken to be solar-system abundances scaled to a metallicity 
of [Fe/H] $=-2.7$. For
elements heavier than the Fe group, we use solar-system
r-process abundances also scaled to [Fe/H] $=-2.7$. 

\begin{figure}[ht]
 \includegraphics[width=0.5\textwidth,clip]{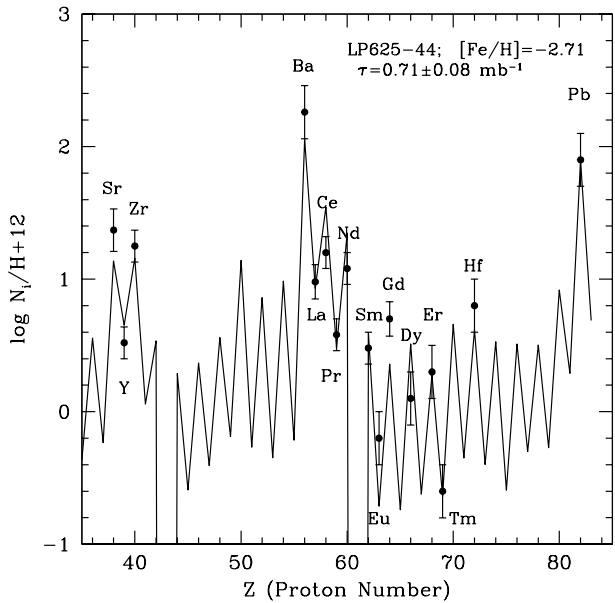}
 \vspace*{-0.7cm}
\caption{Comparison of the observed abundances of LP625-44 with
 the best-fit model result with a
 neutron exposure per pulse of $\tau=0.71\pm 0.08$~mb$^{-1}$.}
   \label{fig1}
\end{figure}

\begin{figure}[ht]
 \includegraphics[width=0.5\textwidth,clip]{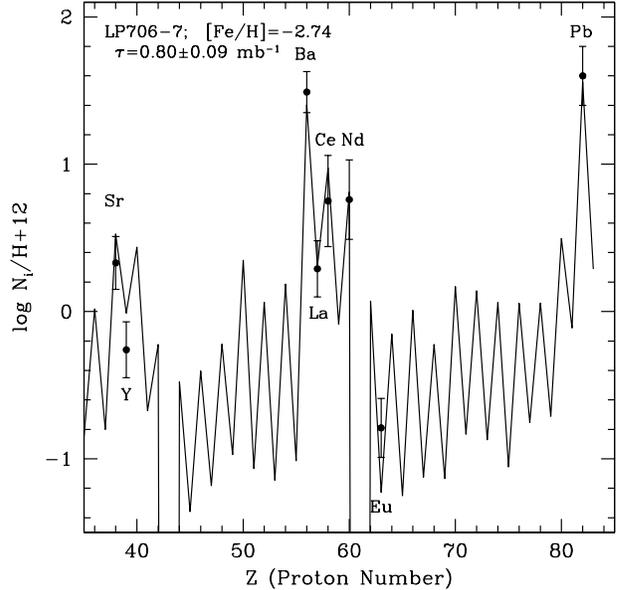}
 \vspace*{-0.7cm}
\caption{Comparison of the observed abundances of LP706-7 with
 the best-fit model result with a neutron exposure per pulse of
 $\tau=0.80\pm 0.09$~mb$^{-1}$.}
 \label{fig2}
\end{figure}

{\bf Figures~\ref{fig1}} and~{\bf \ref{fig2}} show the best fit results for
LP625-44 and  
LP706-7, respectively. The {\it s}-process environment for these models is
characterized by an overlap factor of  $r=0.1$, where $r$ is a fraction
of material from one 
neutron irradiation which survives to the next neutron irradiation.
The best fit neutron exposures per pulse are  $\tau=0.71\pm 0.08$ 
and $0.80\pm 0.09\ {\rm mb}^{-1}$ for LP625-44 and LP706-7, respectively,
corresponding to mean neutron exposures $(\tau_0 = -\tau/\ln{r}$) of
$\tau_0 = (0.58 \pm 0.06) (T_8/3.48)^{1/2}$ and $(0.65 \pm 0.07) (T_8/3.48)^{1/2}$,
respectively.

Such neutron exposures are quite high compared to parameters characterizing 
solar-system abundances for which the main $s$-process component is fit
with $\tau_0 = (0.30 \pm 0.01) (T_8/3.48)^{1/2}$.  Achieving such a high
mean neutron exposure implies that
very few thermal pulses could have contributed to make the {\it s}-process
abundances for these two stars. In fact, in our model
calculations, almost all elements except for
Pb, are made in the first neutron irradiation with the adopted
exposure. Although Pb abundance is more sensitive to the
number of pulses, it too converges to its equilibrium abundance after 
only a few episodes.
Indeed, a nearly equivalent fit to these data can be made in a model
with only a single neutron exposure.  
 However, one can not distinguish
from the mean neutron exposure value whether the thermal pulse or interpulse
phase site is more viable.
This is because 
there is a degeneracy in neutron number density times irradiation time. 
Either site is equivalent as long as
$N_n \Delta t$ gives the same neutron
exposure (see the definition Eq.~(1)),

\begin{figure}[ht]
\centering
 \includegraphics[width=0.5\textwidth,clip]{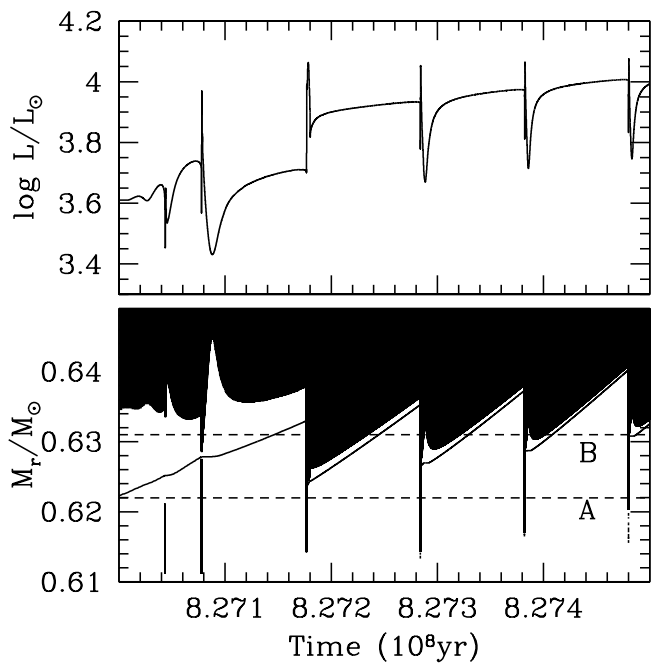}
   \caption{Evolution of a $2M_\odot$ stellar model on the AGB. The upper
  panel shows the variation of surface luminosity 
 accompanying the thermal pulses.  The lower panel shows the evolution of the 
 internal convective zones. The shaded area denotes the envelope
 convective zone and  the vertical lines in the He layer designate the He
  flash convective shell. The thick solid line indicates the H/He
  discontinuity. The two dashed horizontal lines shown by A and B are the mass
  coordinate $M_r/M_\odot=0.622 $ and 0.631, respectively.}
   \label{fig4}

\end{figure}

\subsection{{\it s}-Process in AGB Stars}

These findings are of interest because this is exactly the behavior
anticipated by Fujimoto {\it et al.}~\cite{Fujimoto00}. 
They proposed that only a few 
episodes of proton mixing into the He intershell layer
occur in metal-deficient
stars with [Fe/H] $< -2.5$.  This mixing is invoked by 
an upward extension of the  He convective shell triggered by
a thermal runaway of He shell burning.
Therefore, only a few neutron exposures
can be obtained. The metallicity for LP625-44 and LP706-7 
([Fe/H] $sim -2.7$) enters exactly this metallicity range. It is thus
 possible that the 
{\it s}-process elements observed in these two stars are produced by this 
peculiar mechanism. 

In the present work we have further investigated the time evolution of
the He convective shell.  We have calculated 
the stellar evolution of a  2$M_\odot$ model star with 
[Fe/H] $=-2.7$~\cite{Iwamoto01}. We found that the penetration of
He convective shell into the 
H-rich envelope takes place at the second thermal pulse and that the
convective shell separates into two  
parts.  One is sustained by H-burning and the other is sustained 
by He-burning ({\bf Fig.~\ref{fig4}}). 
The prompt third dredge-up then
follows after the thermal runaway of both H and He
burning terminates. In this way newly
synthesized elements like CNO elements and lithium are brought
into outer convective envelope. Thus, we see these stars as
carbon-rich stars. 
These are usual third dredge-up events.
However, as seen in Fig.~\ref{fig4},
after the first two pulses no more  proton
mixing occurs although third dredge-up events continue
to repeat.

\begin{figure}[ht]
 \begin{center}
 \includegraphics[width=0.5\textwidth,clip]{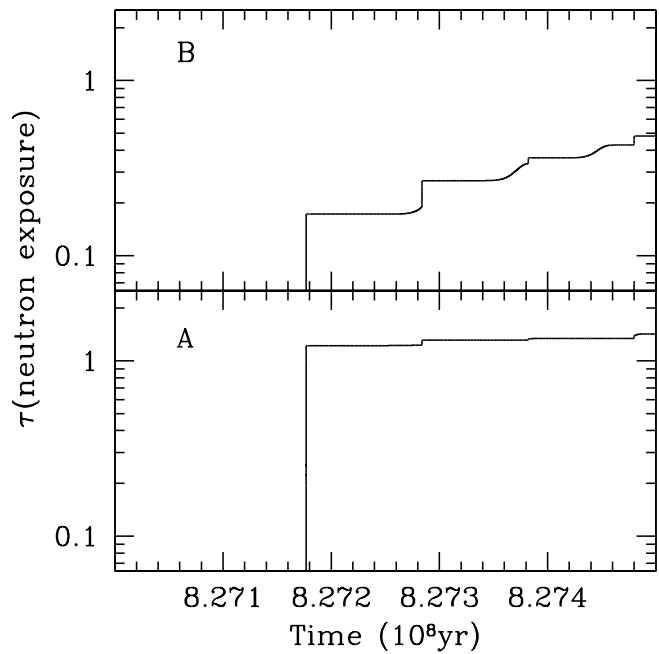}
   \caption{Evolution of neutron exposure at the two mass coordinates A
  and B as  shown in Fig.~\ref{fig4}. }
   \label{fig5}
 \end{center}
\end{figure}

Let us consider when the outward extending He convective shell penetrates
into H-rich outer layer. 
In the He convective shell, the CN cycle operates and produces $^{13}$C and
$^{14}$N from the proton capture on $^{12}$C which has been 
synthesized by the triple
$\alpha$ reaction. Under these circumstances $^{13}$C can
easily capture an alpha particle and produce neutrons via
$^{13}$C($\alpha$, n)$^{16}$O. 
We have used the neutron exposure evaluated in this way to
estimate the influence of this proton mixing on the
 {\it s}-process.
We used a stellar evolution code which included all
neutron capture reactions up to
Si isotopes.  Note that we
do not introduce protons into He layer artificially (i.e., no $^{13}$C
pocket is assumed).
The resultant neutron exposure with time  is 
illustrated in {\bf Fig.~\ref{fig5}}. The first sharp rise in neutron
exposure at the second thermal pulse is due to rapid neutron production
by alpha capture on $^{13}$C at both mass coordinates A and B (see
Fig.~\ref{fig4}). Inside the He convective shell (lower panel~A in
Fig.~\ref{fig5}) this neutron exposure dominates over the subsequent
ones. However, at the mass coordinate~B (upper panel in Fig.~\ref{fig5}) 
all successive neutron irradiations have a
significant influence. The neutron production reaction gets activated during
the interpulse phases as well as during thermal pulses, as the number of
thermal pulse cycle increases.

\section{Conclusion and Discussions}

The two metal-deficient stars, LP625-44 and LP706-7, show 
a strong enhancement of
{\it s}-process elements. Therefore, the abundance ratios of
Ba/Sr and Pb/Ba provide  a precious opportunity to understand the {\it s}-process
mechanism in metal-deficient AGB stars.  First, we used a schematic
pulsed {\it s}-process 
model to estimated the neutron exposure per pulse for these stars. We
found neutron exposures $\tau \sim 0.71$ and 0.80 mb$^{-1}$ for LP625-44 and
LP706-7, respectively. We recognized that these abundance ratios can be
explained by high neutron number density as well as low one, which
means that a true site for {\it s}-process, interpulse or thermal pulse,
cannot be distinguished by this schematic model.  Nevertheless,
it seems likely that the implied high mean neutron exposures probably 
indicate that very few pulses contribute to the observed
{\it s}-process abundances.

Secondly, we have calculated the evolution of 
$2M_\odot$ metal-deficient ([Fe/H]
$=-2.7$) stars  up to  the 
AGB phase. We found that these low-mass, metal-deficient stars 
experience proton mixing into the He intershell convection zone.  
Therefore, an H-flash occurs early in the thermally
pulsing AGB phase.  
This brings about a large change in the surface chemical composition. 
Significant amounts of $^{12}$C and $^{13}$C are dredged-up during a
single deep convective-mixing episode.  
The surface composition then changes from being oxygen-rich to carbon-rich.

We have estimated the neutron exposure in the H-flash convective zone
using a detailed stellar evolution model and a reaction network
extending up to Si isotopes~\cite{Iwamoto01}.
We obtained a distribution of exposures for various mass zones
which can be even higher than $\tau \sim 1$ mb$^{-1}$.  Thus the
high neutron exposures necessary to account for
s-process abundances observed in the low-metallicity stars, LP 625-44 and 
LP706-7~\cite{Aoki01}, are easily explained in this model. 

In our stellar evolution models, material experiencing the {\it s}-process in
the H-flash convective zone is dredged-up by the immediate penetration of
the convective envelope. The {\it s}-process occurs in the He-flash convective
shell, too. The neutron exposure in the
He-flash convective shell is higher than the neutron exposure in the
H-flash convective shell. 
The flash-driven convective shell at the next thermal pulse mixes 
the previously processed material within the He layer. 
After the thermal pulse the third dredge-up brings this material to the
surface. Thus, material exposed to high neutron irradiation appears at
the stellar surface. In the interpulse period
$^{13}$C remains to be
burnt. As the evolution proceeds, $^{13}$C is consumed
rapidly in both the interpulse phase and in the next thermal pulse. 
The {\it s}-process in 
these conditions is, thus, somewhat  complicated and we will 
need further investigation to fully understand the detailed abundance profiles
produced in such low-mass, metal-deficient AGB stars.

\section*{{\fontsize{10}{12}\bf Acknowledgment}}

This work has been supported in part by the Grant-in-Aid for Science
Research (10044103, 10640236, 12047233, 13640313) of the Ministry of
Education, Science, Sports, and\newpage\noindent Culture of Japan, and
also in part by DoE Nuclear Theory Grant (DE-FG02-95-ER40394 at UND).


\setlength{\unitlength}{1mm}
\begin{picture}(81,10)
 \thicklines
 \put(-47.5,8){\line(1,0){81}}
\end{picture}


\begin{thebibliography}{99}
\fontsize{9}{11}\rm


\bibitem{Beers92} T.C. Beers, G.W. Preston, S.A. Shectman, 
	``A search for stars of very low metal abundance. II,''
	{\it Astronomical. J.}, {\bf 103}, 1987, (1992).


\bibitem{Hill00} V. Hill, {\it et al.}, 
	``Heavy-element abundances in the CH/CN-strong very metal-poor
	stars CS 22948-27 and CS 29497-34,''
	 {\it Astron. and Astrophys.}, {\bf 353}, 557, (2000).



\bibitem{Norris97} J.E. Norris, S.G. Ryan, \& T.C. Beers,
	``Extremely Metal-poor Stars. IV. The Carbon-rich Objects,''
	{\it Astrophys. J.}, {\bf 488}, 350 (1997). 

\bibitem{Aoki00}
	W. Aoki, J.E. Norris, S.G. Ryan, T.C. Beers, H. Ando, 
	``Detection of Lead in the Carbon-rich, Very Metal-poor Star LP
	625-44: A Strong Constraint on S-Process Nucleosynthesis at Low
	Metallicity,'' 
	{\it Astrophys. J. Lett}, {\bf 536}, L97 (2000).


\bibitem{Aoki01}
	W. Aoki, S.G. Ryan, J.E. Norris, T.C. Beers, 
	H. Ando, N. Iwamoto, T. Kajino, G.J. Mathews, M.Y. Fujimoto,
	``Neutron Capture Elements in s-Process-Rich, Very Metal-Poor
	Stars,''
	{\it Astrophys. J.}, {\bf 561}, 346, (2001).


\bibitem{Straniero95} O. Straniero, R. Gallino, M. Busso, A. Chieffi,
	C.M. Raiteri, M. Limongi, M. Salaris,
	``Radiative C-13 burning in asymptotic giant branch stars and
	s-processing,'' 
	{\it Astrophys. J. Lett.}, {\bf 440}, L85, (1995).

\bibitem{Busso99} 
	M. Busso, R. Gallino, G.J. Wasserburg,
	``Nucleosynthesis in Asymptotic Giant Branch Stars: Relevance for
	Galactic Enrichment and Solar System Formation,''
	{\it Ann. Rev. Astron. and Astrophys.}, {\bf 37}, 239,
	(1999).

\bibitem{Ryan00} S.G. Ryan, W. Aoki, L.A.J. Blake, J.E. Norris, 
	T.C. Beers, R. Gallino, M. Busso, H. Ando, 
	``s- and r-process elements in two very metal-poor stars,''
	astro-ph/0008423, (2000). 


\bibitem{Fujimoto00} M.Y. Fujimoto, Y. Ikeda, I. Iben, Jr.,
	``The Origin of Extremely Metal-poor Carbon Stars and the Search
	for Population III,''
	{\it Astrophys. J. Lett.}, {\bf 529}, L25, (2000).


\bibitem{vaneck01} S. Van Eck, S. Goriely, A. Jorissen, B. Plez, 
	``Discovery of three lead-rich stars,''
	{\it Nature}, {\bf 412}, 793 (2001).

\bibitem{Mowlavi00} S. Goriely, N. Mowlavi, 
	``Neutron-capture nucleosynthesis in AGB stars,''
	{\it Astron. \& Astrophys.}, {\bf 362}, 599, (2000). 

\bibitem{Howard86}
	W.M. Howard, G.J. Mathews, K. Takahashi, R.A. Ward,
	``Parametric study of pulsed neutron source models for the
	s-process,'' 
	{\it Astrophys. J.}, {\bf 309}, 633 (1986).


\bibitem{Bao00} Z.Y. Bao, {\it et al.}, 
	``Neutron Cross Sections for Nucleosynthesis Studies, ''
	{\it Atomic Data and Nuclear Data Tables}, {\bf 76}, 70, (2000).


\bibitem{Iwamoto01} N. Iwamoto, T. Kajino, M.Y. Fujimoto, G.J. Mathews,
	W. Aoki, ``Flash-Driven Convective Mixing in Low-Mass,
	Metal-Deficient AGB Stars: A New Paradigm for Lithium and
	s-Process Enrichment,'' submitted to {\it
	Astrophys. J.} (2002).


\end{thebibliography}
\end{document}